# Field induced density wave in a kagome superconductor

**Authors:** Md Shafayat Hossain[1]*†, Qi Zhang[1]*, Julian Ingham[2]*, Jinjin Liu[3,4,5]*, Sen Shao[6], Yangmu Li[7], Yuxin Wang[8], Bal K. Pokharel[8], Zi-Jia Cheng[1], Yu-Xiao Jiang[1], Maksim Litskevich[1], Byunghoon Kim[1], Xian Yang[1], Yongkai Li[3,4,5], Tyler A. Cochran[1], Yugui Yao[3,4], Dragana Popović[8], Zhiwei Wang[3,4,5]†, Guoqing Chang[6], Ronny Thomale[9], Luis Balicas[8], M. Zahid Hasan[1]†

**Affiliations:**

[1] Laboratory for Topological Quantum Matter and Advanced Spectroscopy, Department of Physics, Princeton University, Princeton, New Jersey, USA.

[2] Department of Physics, Columbia University, New York, New York 10027, USA.

[3] Centre for Quantum Physics, Key Laboratory of Advanced Optoelectronic, Quantum Architecture and Measurement (MOE), School of Physics, Beijing Institute of Technology, Beijing 100081, China.

[4] Beijing Key Lab of Nanophotonics and Ultrafine Optoelectronic Systems, Beijing Institute of Technology, Beijing 100081, China.

[5] Material Science Center, Yangtze Delta Region Academy of Beijing Institute of Technology, Jiaxing 314011, China.

[6] Division of Physics and Applied Physics, School of Physical and Mathematical Sciences, Nanyang Technological University, 21 Nanyang Link, 637371, Singapore

[7] Beijing National Laboratory for Condensed Matter Physics Institute of Physics, Chinese Academy of Sciences Beijing, 100190, P.R. China

[8] National High Magnetic Field Laboratory, Tallahassee, Florida 32310, USA.

[9] University of Würzburg, Am Hubland 97074 Würzburg, Deutschland.

†Corresponding authors, E-mail: mdsh@princeton.edu; zhiweiwang@bit.edu.cn; mzhasan@princeton.edu.

*These authors contributed equally to this work.

**Abstract:**

On the kagome lattice, electrons benefit from the simultaneous presence of band topology, flat electronic bands, and van Hove singularities, forming competing or cooperating orders. Understanding the interrelation between these distinct order parameters remains a significant challenge, leaving much of the associated physics unexplored. In the kagome superconductor $KV_3Sb_5$, which exhibits a charge density wave (CDW) state below $T \simeq 78$ K, we uncover an unpredicted field-induced phase transition below 6 K. The observed transition is marked by a hysteretic anomaly in the resistivity, nonlinear electrical transport, and a change in the symmetry of the electronic response as probed via the angular dependence of the magnetoresistivity. These observations surprisingly suggest the emergence of an unanticipated broken symmetry state coexisting with the original CDW. To understand this experimental observation, we developed a theoretical minimal model for the normal state inside the high-temperature parent CDW phase where an incommensurate CDW order emerges as an instability sub-leading to superconductivity. The incommensurate CDW emerges when superconducting fluctuations become fully suppressed by large magnetic fields. Our results suggest that, in kagome superconductors, quantum states can either coexist or are nearly degenerate in energy, indicating that these are rich platforms to expose new correlated phenomena.

**Main Text:**

Strongly correlated materials harbor a plethora of competing ground states (*e.g.*, magnetism, density waves, and superconductivity). Comprehending these competing/coexisting quantum phases is a significant challenge in quantum materials. Kagome lattices can act as new windows for understanding the correlation-driven quantum phases. They encompass multiple electronic features like van Hove singularities and flat bands, which can lead to intertwined correlations and exotic quantum states [1]. In the kagome lattice superconductors $A$V$_3$Sb$_5$ [2], where $A$ represents alkali metals K, Rb, or Cs, below the charge density wave (CDW) transition temperature ($T_{\text{CDW}}$ = 79 K – 102 K depending on the specific alkali element), an unconventional CDW emerges due to such an interplay [3-12]. This transition prompts the vanadium atoms to self-reorganize in a star-of-David pattern within the kagome plane. These Kagome systems exhibit intricate three-dimensional CDW arrangements depending on the alkali element. X-ray scattering experiments have revealed superstructures like $2 \times 2 \times 1$ and $2 \times 2 \times 2$, or a combination of both [7]. Additionally, scanning birefringence microscopy [8] and scanning tunneling microscopy (STM) [9] studies have hinted at the breaking of the six-fold rotation symmetry and the presence of nematicity in the charge ordered state. Moreover, it has been observed that the CDW coexists with superconductivity at lower temperatures [10], suggesting pair density wave formation. Thus, the CDW state in AV$_3$Sb$_5$ has remained a notable source of intrigue since its discovery. Importantly, the major excitements in the CDW state of these kagome lattices came in the context of time-reversal symmetry breaking, leading to a very exotic CDW state, suggesting the existence of long-debated loop currents akin to what was suggested for cuprates [13-15]. Experimental techniques such as STM [3-5], muon spin relaxation [6], optical Kerr [8], and electrical transport [12] measurements have provided evidence for the spontaneous breaking of time-reversal symmetry within the charge order. However, contradictory results have also been reported [9, 12, 16] with no clear evidence for magnetic order in the AV$_3$Sb$_5$ systems [17]. In the absence of a magnetic order, one would not expect a sudden change in a material's transport response as a function of magnetic field. In that context, we uncover a surprising and unpredicted phenomenon in KV$_3$Sb$_5$: a field-induced broken symmetry state that emerges at a first-order phase transition at high magnetic fields close to 30 T.

To unravel the transport response of KV$_3$Sb$_5$ with respect to high magnetic fields, we fabricated devices containing mechanically exfoliated KV$_3$Sb$_5$ flakes, which are approximately 100 nm thick, as depicted in Fig. 1(a). Figure 1(b) illustrates the temperature-dependent resistance ($R_{\text{xx}}$) of the sample, revealing metallic transport and a transport anomaly at approximately $T = 77$ K, marking the CDW transition [2], in addition to the superconducting transition reported below 1 K [18]. Having confirmed that our sample displays the hallmarks of the correlated phases in KV$_3$Sb$_5$, we move on to explore high-field electrical transport. Figure 1(c), which depicts $R_{\text{xx}}$ as a function of the magnetic field ($\mu_0H$) applied along the sample's crystallographic $ab$ plane and perpendicular to the direction of the current, acquired at $T = 0.56$ K, showcases a peculiar transition. $R_{xx}$ displays a sharp increase in slope, *i.e.*, $dR_{xx}/d\mu_0H$, between ~27-32 T, with similar low-field and high-field slopes. Notably, the onset of this sharp slope change depends on the sweeping direction of $\mu_0H$ [Fig. 1(c)], exhibiting hysteresis as a function of $\mu_0H$, indicating a first-order phase transition.

Upon observing the phase transition, we investigated its temperature dependence. Figure 1(d) illustrates a series of magnetoresistance traces obtained at temperatures ranging from 0.56 K to 12 K. At the lowest temperatures, or for $T$ between 0.56 K and 1.5 K, the transition near $\mu_0H = 30$ T is quite evident. However, its signature weakens considerably at 4 K and becomes nearly indiscernible at higher temperatures. To visualize the temperature dependence of the transition, we first define the critical magnetic field for the transition ($\mu_0H_c$) as the magnetic field value where $dR_{xx}/d\mu_0H$ reaches its maximum. Subsequently, we plot $dR_{xx}/d\mu_0H$ at $\mu_0H_c$ as a function of the temperature, as shown in the inset of Fig. 1(d). This plot reveals a distinct temperature dependence, allowing us to define the onset of the transition around $T = 6$ K.

After analyzing the temperature dependence of the transition, we investigated how this transition affects other transport properties of the $KV_3Sb_5$ sample. We conducted measurements of $R_{xx}$ as a function of a rotating in-plane magnetic field to understand the symmetry of the transport response, as shown in Fig. 2(a). Such measurements can reveal the symmetry of the angle-dependent magnetoresistivity, providing insights into the underlying transport characteristics. $R_{xx}(\theta)$, with $\theta$ being the azimuthal angle defining the in-plane rotations concerning the original direction of the electrical current, displays nearly four-fold symmetric oscillations at low fields [see Figs. 2(b-d)]. Due to the Lorentz force, one should expect two-fold symmetric angular oscillations, with a minimum in the resistance for fields along the current and a maximum for currents perpendicular to it (maximum interaction with the field). Notably, this four-fold symmetry survives the application of fields as intense as $B = 28$ T, indicating that the four-fold symmetry, or four-fold symmetric carrier scattering, is intrinsic to the Fermi surface of the CDW ground state for reasons that remain to be understood [19]. This observation of near four-fold symmetry aligns with previous findings [19]. Figure S1 showcases the angular magnetotransport data collected under $B = 18$ T at different temperatures, indicating that the angular magnetoresistance is isotropic above the CDW transition temperature, $T_{CDW} = 77$ K [Fig. 1(b)]. At these higher temperatures, the charge carriers are predominantly scattered by phonons. Consequently, we conclude that the transport only exhibits a four-fold symmetry when the CDW state develops.

Having observed the four-fold symmetric $R_{xx}(\theta)$ at magnetic fields below $\mu_0H_c$, we proceed to apply magnetic fields above $\mu_0H_c$. Interestingly, we find that for fields beyond $\mu_0H = 30$ T, $R_{xx}(\theta)$ becomes entirely two-fold symmetric [Figs. 2(g, h)], as one would expect from the Lorentz force argument (discussed above). This implies an electronic reconstruction of the Fermi surface either towards a more symmetric Fermi surface or a Fermi surface characterized by a nearly uniform scattering rate as a function of the azimuthal angle. Therefore, the change in the angular dependence of the magnetoresistance with increasing magnetic field supports the notion of a magnetic field induced phase transition.

Next, we compare the temperature dependence of the angular magnetoresistance for fields below (18 T) and above (40 T) the field-induced transition. Figure 3 illustrates the results of these experiments. In Fig. 3(a), we present the data collected under 18 T, showcasing angular magnetoresistance traces as a function of the in-plane magnetic field orientation for temperatures ranging from 0.56 K to 8.0 K. Across this temperature range, the data consistently shows four-fold symmetry. Conversely, at 40 T, which surpasses the magnetic field onset of the field-induced transition, we observe a two-fold symmetric angular magnetoresistance at low temperatures, specifically 0.56 K and 1.5 K [Fig. 3(b)]. However, as the temperature increases, for instance, to 8.0 K, the angular magnetoresistance reverts to a four-fold symmetric pattern. Notice that $T = 8.0$ K is above the critical temperature required to observe the field-induced transition. Therefore, this observation indicates that above the field-induced transition critical temperature, the transport response at high fields becomes akin to that observed at low fields. Thus, we conclude that the two-fold symmetric angular magnetoresistance is a consequence of the field-induced transition. It is worth noting that the angular dependence of the critical magnetic field of the transition, $\mu_0 H_c$ itself exhibits an emergent two-fold symmetry (Fig. S2) reminiscent of the two-fold symmetric superconducting transitions observed in several transition metal dichalcogenides [20, 21].

The carrier scattering rate on the Fermi surface defines the angular dependence of the magnetoresistivity. Therefore, any change in the symmetry of the angular magnetoresistivity points to a modification in the geometry of the original Fermi surface at the field-induced transition, which in turn affects the scattering rate. As we discuss below, a magnetic field induced CDW state is likely to be associated with those Fermi surface sheets responsible for the superconductivity at zero field.

To explore the possibility of a CDW emerging during the field induced transition, we investigated the current-voltage characteristics of the system both before and after the transition. Figure 4 summarizes the outcome of our investigation. At $\mu_0 H = 40$ T, which exceeds $\mu_0 H_c$, we observe pronounced nonlinear differential resistance with a characteristic threshold voltage [Fig. 4(a)], often associated with the sliding phason mode of a CDW phase [22-24]. Specifically, incommensurate CDWs are typically pinned by the point disorder. When exposed to a sufficiently large electric field, the pinning force is overcome, leading to CDW depinning and sliding, resulting in a sharp decrease in resistance [22-24]. Our analysis reveals that the depinning threshold electric field ($E_c = R I_c / l$; where $I_c$ is the critical current for depinning and $l$ is the sample length, i.e., the separation between the current contacts) is $\sim 11$ mV/cm. Such $E_c$ values of the order of 10 mV/cm are commonly observed in incommensurate CDW systems such as $NbSe_3$ and $TaS_3$ [22, 23]. Furthermore, when we raised the temperature to 8 K, which surpasses the critical temperature for the field-induced transition, the $I$-$V$ characteristics became linear (Ohmic-like), indicating that the nonlinear response is intrinsic to the field-induced transition.

Additionally, in Fig. 4(b), we compare the differential resistance obtained prior to transition (18 T) and after the transition (40 T) at $T = 0.56$ K. While the 40 T trace exhibits pronounced nonlinearity, the 18 T trace at the same temperature displays linear response. This substantiates that the nonlinear response indeed arises from the field-induced transition. Considering that the ground state prior to the field-induced transition is a CDW state, one would expect also to observe nonlinearity in the transport. However, the CDW ground state before the field-induced transition is a commensurate 2×2×1, 2×2×2, or a combination of both [7], which is strongly pinned to the lattice and would require a very high electric field to depin it. Consequently, one would anticipate a significantly larger depinning threshold electric field (*e.g.*, 1 kV/cm [25]) than the ~11 mV/cm observed in the field-induced state. Overall, the observed nonlinear response suggests the development of an incommensurate CDW at the field-induced transition.

To understand this surprising field-induced phase transition, we developed a minimal theoretical model. Field-induced charge orders have been observed in cuprates [26, 27], marking an intriguing parallel with other similarities observed between cuprates and KV$_3$Sb$_5$, such as the possible existence of loop currents [13], reminiscent of the Varma model for cuprates [14, 15]. Consonant with theories proposed in the context of cuprates [28], we suggest that the observed transition in the magnetoresistance is due to a CDW which, at lower fields, is suppressed by a dominant superconducting phase. Other measurements in the AV$_3$Sb$_5$ kagome family have observed evidence for an extended window of superconducting fluctuations above $T_c$ [29], and high fields may thus be necessary to fully resolve the competitive or intertwined nature of a subleading charge density wave and superconductivity. To provide a microscopic understanding of this scenario, we briefly discuss the electronic structure at low temperatures and zero field, illustrated in Fig. 5. After the onset of the parent CDW order near $T = 77$ K, a set of three hole-like pockets appear inside the reconstructed band structure [30] primarily composed of the vanadium $d$-orbitals. These pockets appear centered near the $M$ point of the mini-Brillouin zone and at finite $k_z \sim 0.55\pi/c$, and exhibit a hole-like dispersion $E_V \sim \mu_V - \frac{1}{2m_V}k^2$. Largely unaffected by the Fermi surface reconstruction is the electron-like pocket at the $\Gamma$ point deriving from the Sb $p_z$ orbitals, with a dispersion $E_{Sb} \sim \mu_{Sb} + \frac{1}{2m_{Sb}}k^2$ [31]. The presence of electron-like and hole-like pockets near the Fermi energy results in an instability towards orbital hybridization known as excitonic order [32]. The difference in areas of the two Fermi surfaces, however, suppresses this instability, making way for superconducting order to develop at zero field on the Fermi surface with the largest density of states, *i.e.*, the electron pocket at the $\Gamma$-point.

In the Supplemental Material, we analyze a minimal interacting model of these two Fermi surfaces and find a CDW order that hybridizes the bands associated with the Sb and V orbitals for repulsive interactions. The result is a finite-momentum excitonic insulator, producing a CDW that is incommensurate along the $z$-direction as the hole-like pockets are centered at finite $k_z$. The anisotropic spectral weight observed on the folded side of the hole pockets [30] could also produce additional weak in-plane incommensuration due to preferential coupling to 'hot spots' – this could explain the observation of a small threshold electric field needed to depin the CDW (Fig. 4). This mechanism has also been proposed to explain the incommensurate charge density wave seen under pressure or doping in transition metal dichalcogenide TiSe$_2$ – a material which intriguingly exhibits a very similar Fermi surface structure to KV$_3$Sb$_5$ within the CDW phase [33, 34], and has also been shown to exhibit chiral charge order via measurement of the photogalvanic effect [35], and in which the competition with an incommensurate CDW order appears to be closely linked to superconductivity [36, 37]. It is also possible that the large magnetic shifts the bands in a way that further promotes nesting between electron and hole pockets, leaving prospects for future theoretical modeling. We also cannot rule out the possibility of a field-induced alteration of the original CDW (at $\mu_0H = 0$ T) order, which would require structural analyses (*e.g.*, X-ray scattering) under high fields.

It is surprising that despite the differences in electronic structures among KV$_3$Sb$_5$, heavy fermion superconductor CeRhIn$_5$ [38], high-temperature cuprate superconductor YBa$_2$Cu$_3$O$_{7-\delta}$ [26, 27], and graphite [39-41], they exhibit magnetic field-induced density wave orders. Furthermore, theoretically, a magnetic field induced CDW could explain [42] the observation of a three-dimensional (3D) quantum Hall effect (QHE) in ZrTe$_5$ [43]. In $k$-space, the development of a CDW would open a gap in a one-dimensional Landau band, leading to an insulating bulk. In real space, the development of the CDW can split the 3D electron gas into decoupled 2D quantum Hall layers, leading to a 3D QHE regime [42]. Although electron-phonon can be a possible mechanism driving the CDW and 3D QHE, given the array of electronic degrees of freedom (orbital, spin, and valley), one can also expect density waves to arise from electronic correlations or even from exciton condensation [40, 41, 44-47]. Our comprehensive transport measurements indicate that, in $KV_3Sb_5$, an additional broken symmetry state emerges, coexisting with its original CDW, leading to a hitherto unreported magnetic field-induced first-order phase transition to an incommensurate CDW state. An intriguing possibility to be further explored would be the possible existence of additional magnetic field induced CDWs in $KV_3Sb_5$ at even higher magnetic fields, and if these would affect the Hall effect, *i.e.*, leading to a 3D QHE akin to the one proposed for other layered quantum materials like $HfTe_5$, $TaS_2$, or $NbSe_3$ [42]. Our discovery thus not only motivates a battery of experiments under high magnetic fields (such as high-field nuclear magnetic resonance or X-ray diffraction experiments) to determine the nature of the pertinent electronic order and associated wavevectors but also encourages the exploration of additional magnetic field-induced states in kagome compounds under high magnetic fields.

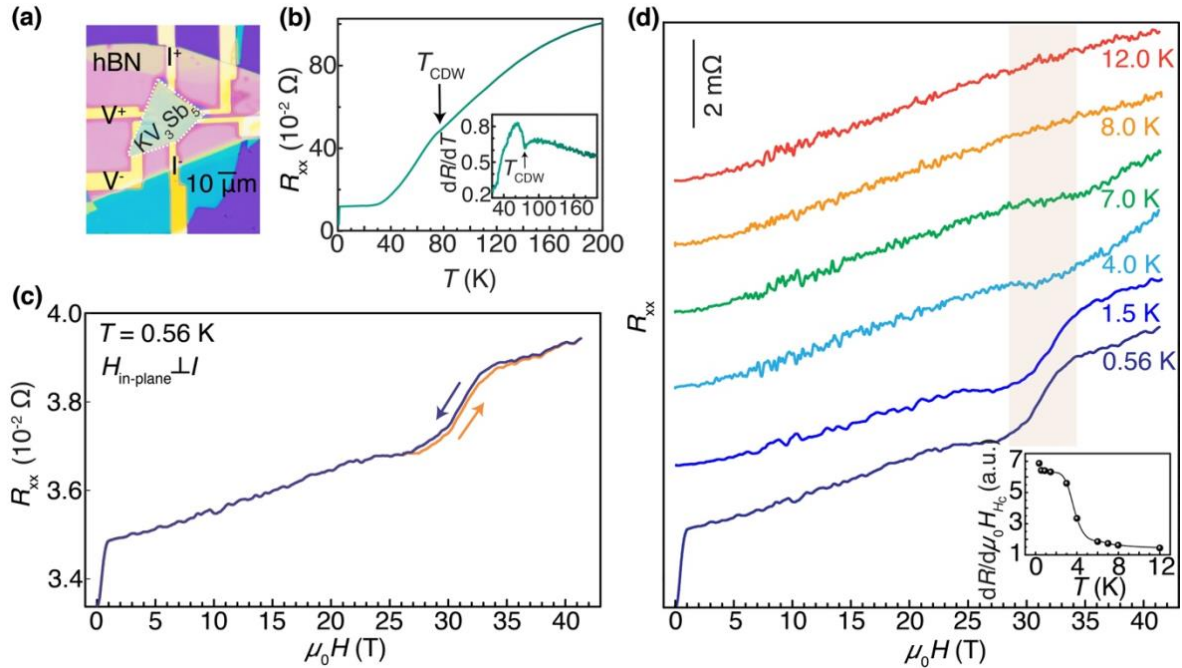

**Fig. 1: Magnetic field-induced transition in KV$_3$Sb$_5$.** (a) Optical microscopy image of the device used for the transport measurements, comprising a mechanically exfoliated KV$_3$Sb$_5$ flake. The white dashed lines indicate the boundaries of the flake. Current (I$^+$ and I$^-$) and voltage (V$^+$ and V$^-$) contacts for the four-probe transport measurements are also indicated. (b) Temperature-dependent longitudinal resistance ($R_{xx}$), exhibiting an anomaly around $T = 77$ K indicating a CDW transition, alongside a sharp decrease in $R_{xx}$ at low temperatures due to the onset of superconductivity. The inset illustrates the derivative of the $R_{xx}$ as a function of $T$, highlighting the transport anomaly centered at the CDW transition near 77 K. (c) $R_{xx}$ as a function of the magnetic field ($B$), with $B$ applied along the sample's in-plane direction but perpendicular to the current. Orange and blue curves represent traces collected during field sweep from 0 to 40 T (up sweep) and 40 T to 0 T (down sweep), respectively. A field-induced transition near 30 T is observed, exhibiting hysteresis with the field sweep direction. (d) Temperature dependence of the field-induced transition, as observed through a series of $R_{xx}$ traces at temperatures ranging from 0.56 K to 12 K. At $T = 8$ K and 12 K, no transition is observed. Inset: derivative of $R_{xx}$ with respect to $B$ (d$R_{xx}$/d($\mu_0 H$)) at the $\mu_0 H$ value where d$R_{xx}$/d($\mu_0 H$) reaches its maximum, plotted as a function of the temperature. The $\mu_0 H$ value where d$R_{xx}$/d($\mu_0 H$) peaks defines the critical field, $\mu_0 H_c$ for the field-induced transition. Thus, the temperature dependence of d$R_{xx}$/d$B$ at $B_c$ reveals the critical temperature for the field-induced transition.

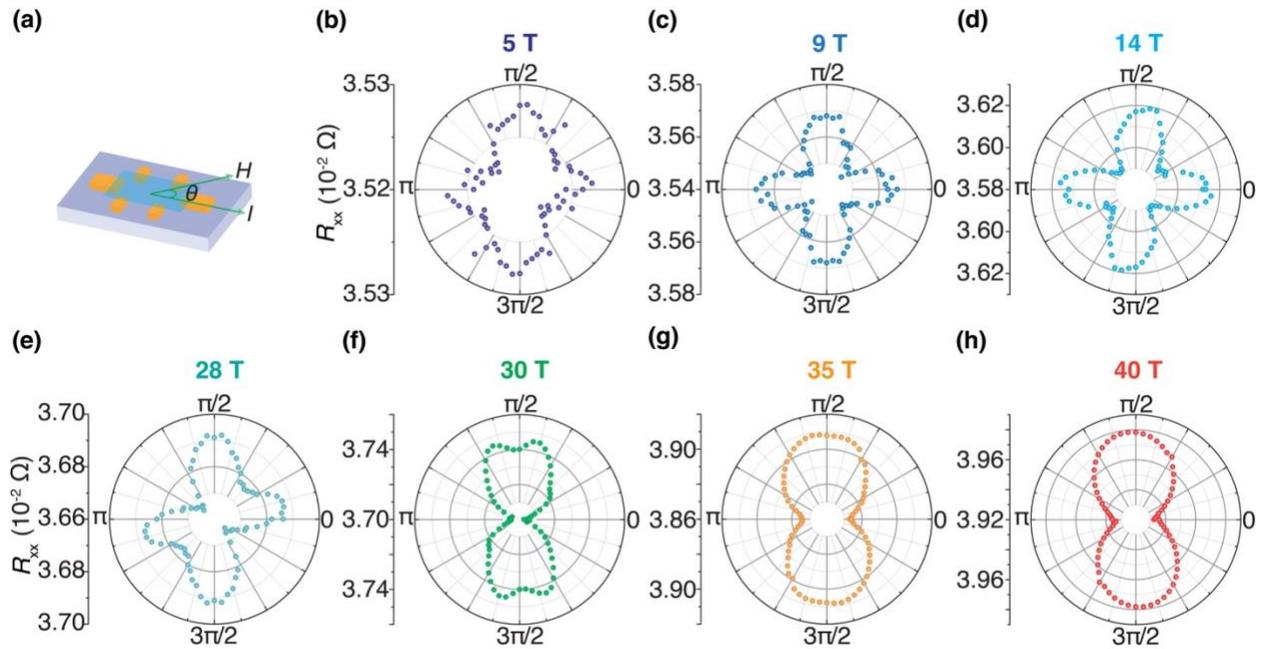

**Fig. 2: Angular dependence of the magnetoresistance across the field-induced transition.** (a) Schematics illustrating the orientation of the in-plane rotating magnetic field, where $\theta$ represents the angle of the magnetic field relative to the current flow direction. (b-h) $\theta$-dependence of $R_{xx}$ measured at $T = 0.56$ K under various magnetic fields ranging from 5 T to 40 T. Prior to the field-induced transition, $R_{xx}$ demonstrates an emerging four-fold symmetric behavior (panels b-d). The four-fold angular dependent magnetoresistance gradually transitions into a two-fold symmetric pattern at the onset of the field-induced transition (e, f) becoming purely two-fold symmetric at higher fields (g, h). The data were collected across the angular range from $\theta = 0$ to $\pi$ and then repeated with $\pi$ periodicity.

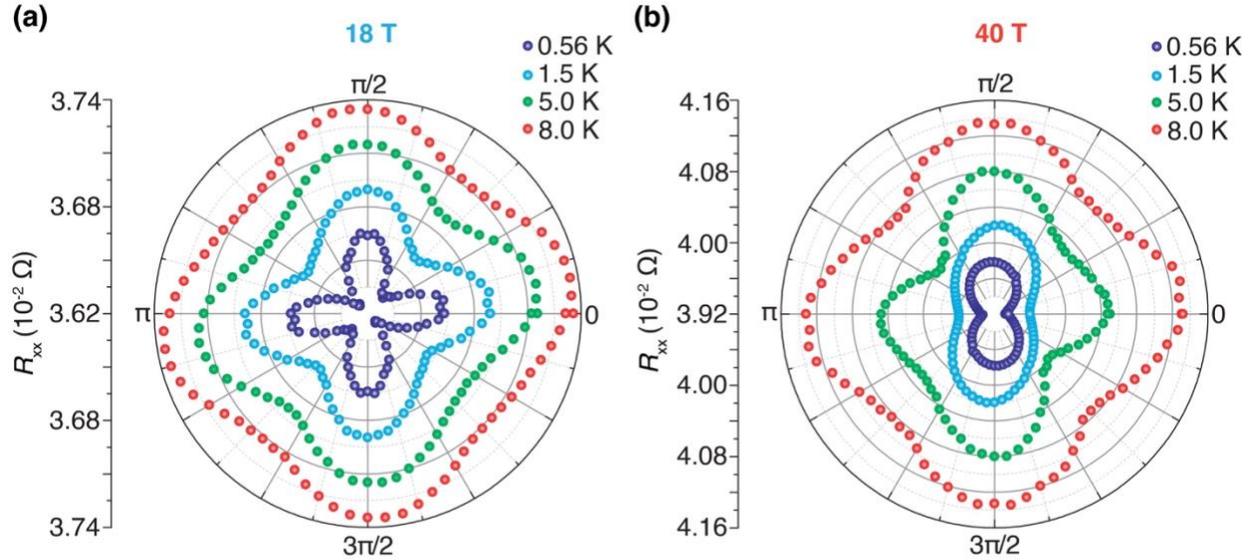

**Fig. 3: In-plane magnetoresistance as a function of the azimuthal angle $\theta$ for fields below and above the field-induced transition and for several temperatures.** (a) Angular dependence of $R_{xx}$ collected under 18 T (below the field induced transition), for several temperatures ranging from 0.56 K to 8.0 K. $R_{xx}$ exhibits a consistent four-fold symmetric behavior across all temperatures. (b) Angular dependence of $R_{xx}$ at 40 T (above the field induced-transition), plotted within the same temperature range as in panel (a), displaying two-fold symmetric behavior at low temperatures but transitioning to a four-fold symmetric one at temperatures exceeding 5 K, thus recovering the behavior observed at low fields. This observation is consistent with the transition temperature identified in Fig. 1(d). The data were collected across the angular range from 0 to $\pi$ and then repeated with $\pi$ periodicity.

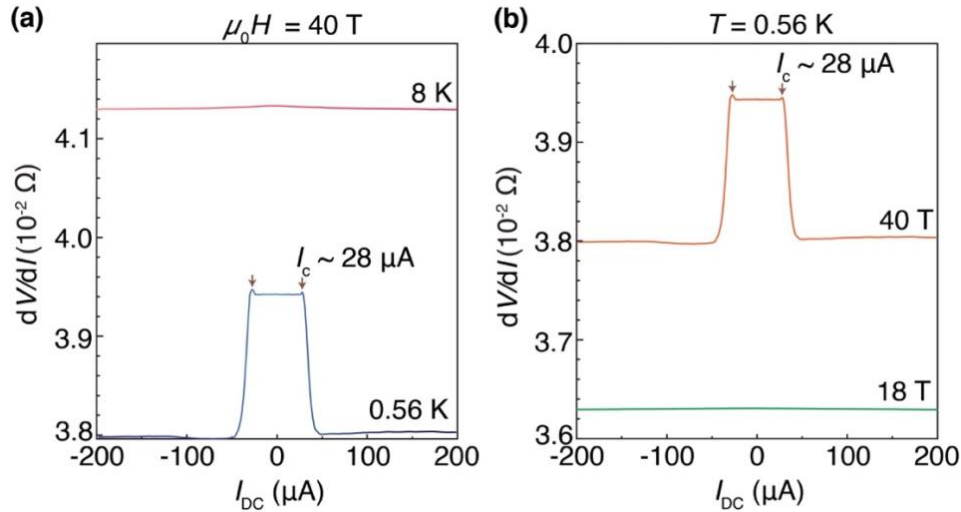

**Fig. 4: Differential in-plane resistance d$V$/d$I$, measured with a small (1 µA) AC current overlaid on a DC current bias, revealing nonlinear transport in the field-induced transition.** (a) d$V$/d$I$ measured under identical conditions at 40 T (with the magnetic field applied at an in-plane angle of 45-degrees with

respect to the current direction) and at two temperatures, 0.56 K and 8 K. At high temperatures (above the critical temperature for the field-induced transition), the system exhibits ohmic behavior over a wide range of current biases, while the low-temperature data displays markedly nonlinear transport, typical of a pinned, incommensurate CDW that is depinned by electric field surpassing the threshold field for depinning. It is worth noting that self-heating would not lead to a well-defined threshold electric field but instead to a parabolic dependence of the d$V$/d$I$ on $I$, in contrast to the observed behavior. (b) d$V$/d$I$ measured under $B = 18$ T (below the field-induced transition) and 40 T (above the transition) with the magnetic field applied at an angle $\theta = 45°$ with respect to the current direction at $T = 0.56$ K. At 40 T, the transport response is nonlinear. The nonlinearity is absent under $B = 18$ T.

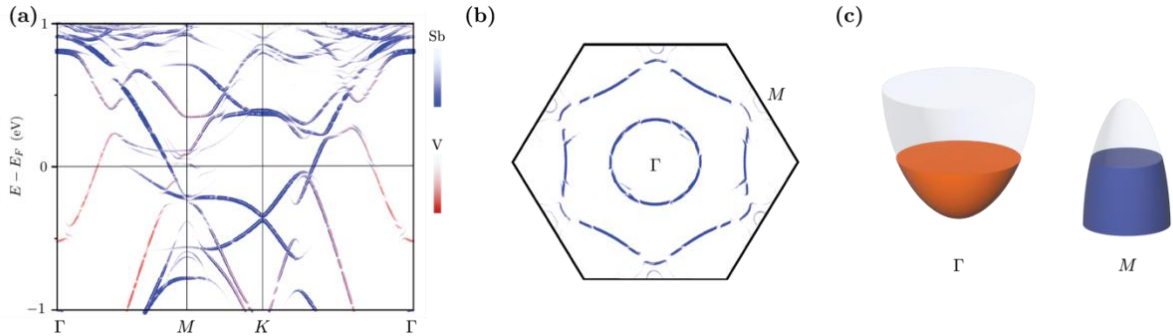

**Fig. 5: Electron and hole pockets in the CDW phase of KV$_3$Sb$_5$.** (a) Band structure of KV$_3$Sb$_5$ within the CDW phase along $k_z = 0$, comprising an electron-like pocket near the Γ-point of the first Brillouin zone, which is primarily composed of the Sb $p_z$ orbitals, and reconstructed portions of the vanadium kagome saddle points near $M$. (b) Unfolded Fermi surface in the Brillouin zone of the pristine phase at $k_z = 0$. Near the $M$-point, one observes hole-like pockets, centered at an incommensurate wavevector $k_z \sim 0.55\pi/c$. (c) Schematic of the hole- (left) and electron-like (right) pockets with different effective masses.

**Acknowledgement:**
We acknowledge illuminating discussions with Titus Neupert. M.Z.H. group acknowledges primary support from the US Department of Energy, Office of Science, National Quantum Information Science Research Centers, Quantum Science Center (at ORNL) and Princeton University; STM Instrumentation support from the Gordon and Betty Moore Foundation (GBMF9461) and the theory work; and support from the US DOE under the Basic Energy Sciences programme (grant number DOE/BES DE-FG-02-05ER46200) for the theory and sample characterization work including ARPES. The sample growth was supported by the National Key Research and Development Program of China (grant nos 2020YFA0308800 and 2022YFA1403400), the National Science Foundation of China (grant no 92065109), and the Beijing Natural Science Foundation (grant nos Z210006 and Z190006). Z.W. thanks the Analysis and Testing Center at BIT for assistance in facility support. D.P. group acknowledges support by National Science Foundation Division of Materials Research through DMR-1707785. L.B. is supported by the US-DoE, Basic Energy Sciences program through award DE-SC0002613. A portion of this work was performed at the National High Magnetic Field Laboratory, which is supported by National Science Foundation Cooperative Agreement No. DMR-2128556 and the State of Florida.